\documentclass[%
 reprint,
superscriptaddress,
amsmath,amssymb,
prb,
]{revtex4-1} 

\usepackage{graphicx}
\usepackage{dcolumn}
\usepackage{bm}

\usepackage{amsmath} 
\usepackage{tikz}

\DeclareRobustCommand{\filledDot}{\tikz\draw[fill=black] (0,0) circle (0.05cm);}
\DeclareRobustCommand{\emptyDot}{\tikz\draw (0,0) circle (0.05cm);}
\DeclareRobustCommand{\smallDot}{\tikz\draw (0,0) circle (0.02cm);}

\DeclareMathAlphabet\mathbfcal{OMS}{cmsy}{b}{n}  

\newcommand*{\citen}[1]{%
  \begingroup
    \romannumeral-`\x 
    \setcitestyle{numbers}%
    \cite{#1}%
  \endgroup   
}

\makeatletter
\let\origsection\section
\renewcommand\section{\@ifstar{\starsection}{\nostarsection}}

\newcommand\nostarsection[1]
{\sectionprelude\origsection{#1}\sectionpostlude}

\newcommand\starsection[1]
{\sectionprelude\origsection*{#1}\sectionpostlude}

\newcommand\sectionprelude{%
  \vspace{-0.8em}
}

\newcommand\sectionpostlude{%
  \vspace{-0.8em}
}
\makeatother

\begin{document}


\title{Linear magnetoelectricity in the Zintl phase \\ 
pnictides $\mathrm{(Ba,Ca,Sr)Mn}_2\mathrm{(P,As,Sb)}_2$ from first principles calculations}


\author{John Mangeri}
 \email{johnma@dtu.dk}
\author{Martin Ovesen}
\author{Thomas Olsen}%
 \email{tolsen@fysik.dtu.dk}
\affiliation{%
Center for Atomic-scale Materials Design (CAMD), Department of Physics, Technical University of Denmark, \\ 
 Anker Engelunds Vej 101 2800 Kongens Lyngby, Denmark\\
}%


\date{\today}

\begin{abstract}

We report a comprehensive set of density functional theory calculations on the family of layered antiferromagnetic manganese pnictides $\mathrm{(Ba,Ca,Sr)Mn}_2\mathrm{(P,As,Sb)}_2$. 
We characterize all components to the linear magnetoelectric (ME) tensor $\alpha$ which are parsed into their contributions from spin and orbital moments for both lattice-mediated and their clamped-ion electronic analogs.
Our main results show that the orbital magnetization components cannot be neglected in these systems.
The ME response is dominated by electronic effects with total $\alpha$ values exceeding those of the prototypical $\mathrm{Cr}_2\mathrm{O}_3$ (i.e. $\alpha \simeq$ 6.79 ps/m in $\mathrm{BaMn}_2\mathrm{As}_2$).
We also identify a strong correlation with the computed ME susceptibility on pnictogen substitution in the trigonal subfamily albeit with weaker amplitudes ($\alpha \approx$ 0.2-1.7 ps/m).
Additionally, we provide the dependence of these predictions on the Hubbard +U correction, at the level of the local density approximation, which show large variations on the calculated ME coefficients in the tetragonal compounds highlighting the role of strong correlation in these compounds.
\end{abstract}


\maketitle


\section{Introduction\label{sec:intro}}

Structures where both spatial inversion ($\mathcal{I}$) and time-reversal ($\mathcal{T}$) symmetry are broken are known as magnetoelectric (ME) materials.
In the atomic picture, the action of an electric field $\mathbfcal{E}$ and/or magnetic field $\mathbfcal{H}$ reorients the nuclear coordinates to a new equilibrium.
The fundamental spin-orbit interaction mediates a \emph{change} of the electron cloud charge center, orientations of the spin magnetization, and corresponding orbital currents relative to that of the ground state in the crystal.
Therefore, the ME effect has contributions from the lattice motion but also a purely electronic signature which includes components from both spin and orbital momentum.
Hence under $\mathbfcal{E}$, changes are expected in the net magnetization $\mathbf{M}$ and the application of an external $\mathbfcal{H}$ drives shifts to the electric polarization $\mathbf{P}$.
Due to this complicated microscopic origin of the effect, it is difficult to make \emph{a priori} estimations of the strength of the ME response based on atomic composition and symmetry alone. 
%
%
To this end, first-principles techniques are indispensable in exploring the different physical pathways involved in inducing  changes in $\mathbf{M}$ and $\mathbf{P}$ by external means.
To first-order, the ME effect can be expressed in the $\{\mathbfcal{E},\mathbfcal{H}\}$ frame in the following form,
\begin{align}\label{eq:linear_ME}
    \alpha_{ij} = \mu_0 \left(\frac{\partial M_i}{\partial \mathcal{E}_j}\right)_{\mathbfcal{H}} = \left(\frac{\partial P_j}{\partial \mathcal{H}_i}\right)_{\mathbfcal{E}}.
\end{align}
The above given in SI units represents the components of a (3$\times$3) pseudotensor ${\bm \alpha}$ with Cartesian indices $i,j$.
A special case is where the magnetic order breaks the inversion symmetry. 
Thus, inversion is coupled to time-reversal, providing $\mathcal{I}\cdot\mathcal{T}$ as a symmetry.
This parity-odd property \cite{Watanabe2018} gives rise to a \emph{linear} term in the ME susceptibility despite formal inversion symmetry of the lattice.
%

%
%

%
Dzyaloshinskii first established interest by predicting the linear ME effect in the antiferromagnetic (AFM) insulator $\mathrm{Cr}_2\mathrm{O}_3$ (CRO) \cite{Dzyaloshinskii1959, Dzyaloshinskii1960}.
A few years later, a series of experiments \cite{Astrov1960, Astrov1961, Folen1961, Rado1961} observed the ME response and since then CRO has become a canonical material to study this phenomena.
A plethora of interesting studies exploiting magnetoelectricity in this compound have been carried out, from large optical excitations \cite{Krichevtsov1993, Krichevtsov1996}, $\mathbfcal{E}$-control of AFM spin-currents \cite{Liu2021}, domain patterns\cite{Hedrich2021}, and surface skyrmions \cite{Du2023}, to strong spin-phonon coupling\cite{Fechner2018} and possible induction of magnetic monopoles near a free surface \cite{Meier2019}.
Using density-functional theory (DFT), the decomposition of lattice and electronic contributions to ${\bm \alpha}$ in CRO has been investigated \cite{Iniguez2008, Bousquet2011, Malashevich2012, Scaramucci2012, Mu2014, Ye2014, Tillack2016, Bousquet2024}.
However, despite more than a decade of effort, precision calculations of CRO's ${\bm \alpha}$ can vary significantly depending on the choice of method, exchange correlation functional, lattice constants, DFT code, and convergence tolerances \cite{Bousquet2024}.
Still, these investigations have uncovered valuable information that the lattice-mediated (LM) and electronic or clamped-ion (CI) spin components dominate the transverse ME response. 
From these studies, the involvement of the $\mathrm{Cr}^{+3}$ orbital moment, which is expected to be strongly quenched in bulk, has been revealed to be negligible (significant) in the tranverse (longitudinal) directions with respect to the AFM ordering direction.
Although CRO continues to be a well-studied benchmark\cite{Bousquet2024}, limited information has been gathered from atomistic methods about other MEs with a few exceptions \cite{Wojdel2009, Yamauchi2010, Scaramucci2012, Ghosh2015, Ricci2016, Dasa2019, Liu2020, Bayaraa2021, Chang2023, Dey2024}.
%
%
%

%
Large ME effects are advantageous from a technological perspective\cite{Liang2021, Kopyl2021} and the search for new MEs is on-going.
Parenthetically, this includes materials that display nonlinear or multiferroic ME coupling, which are \emph{not} the subject of this work, see Refs.~[\citen{Fiebig2016},\citen{Spaldin2019}] for reviews.
We should note that most experimentally observed amplitudes of $\alpha_{ij}$ in different compounds throughout the past two decades are of comparable order to those of CRO \cite{Vaknin2004, Mufti2011, Saha2016, Ghara2017, Yanda2019, Shahee2023, Fogh2023a, Fogh2023b}.
%
%
%
%
However this isn't always the case.
In the ferrimagnetic phase of $\mathrm{Fe}_2\mathrm{Mo}_3\mathrm{O}_8$, one component of ${\bm \alpha}$ was measured\cite{Chang2023} to reach 480 ps/m nearly two orders of magnitude larger although this is only observed above a critical field of 32 T where the material exhibits a phase transition. 
Combining models\cite{Eremin2022} and DFT calculations\cite{Chang2023} have revealed strong exchange-striction, considerable electronic contributions to $\mathbf{P}$, and a large orbital magnetization possibly underpinning the colossal response.
Still, it remains to be seen if other materials can exhibit similar giant linear ME susceptibilities - preferably without the need of inducing a phase transition by means of large external fields.
The new paradigm of materials-by-design provides a path forward in this regard.
However, in order to engineer strongly ME materials, one needs to understand the complex admixing of various physical factors including the spin-orbit interaction, magnetic network geometry and their ligand environments, exchange couplings and anisotropy, orbital hybridization, and the role of electron-electron correlations among others.
To what extent do they influence ${\bm \alpha}$?
We aim to explore aspects of this question in this paper by looking at a specific class of compounds.
%

%
%
Central to simplifying this decompositional analysis, are materials families with closely related structure and electron behavior.
One such family, the AFM lithium orthophosphates, with chemical composition $\mathrm{LiXPO}_4, \mathrm{X} = (\mathrm{Co,Mn,Ni,Fe})$, serve as useful datasets to ascertain how changing the transition metal ion can drastically alter the magnetic anisotropy \cite{Fogh2023a, Fogh2023b}, break the collinear or commensurate spin order \cite{Vaknin2004, Toft-Petersen2015}, lead to ferrotoroidicity\cite{VanAken2007,VanAken2008} or give rise to a pronounced orbital moment response \cite{Scaramucci2012, Yuen2017} thus changing ${\bm \alpha}$.
The Zintl phase layered AFM pnictides with formula $\mathrm{AMn}_2\mathrm{Pn}_2$ with $\mathrm{A} = \mathrm{(Ba, Ca, Sr)}$ and pnictogens $\mathrm{Pn} = \mathrm{(P, As, Sb)}$ comprises another set of such materials that may serve to decouple possible trends in ME properties.
They have been identified as possible candidates for hosting linear ME coupling \cite{Watanabe2018, Lovesey2018} and a wealth of properties have been unravelled experimentally\cite{Bobev2006, Bridges2009, Johnston2011, Simonson2012, Sangeetha2016, Zhang2016, Lovesey2018, Sangeetha2018, Sangeetha2021, Jacobs2023}.
The substitution of the cation from $\mathrm{Ba}$ to $\mathrm{Ca}$ (or $\mathrm{Sr}$) drives a change in unit cell symmetry while the pnictogens alter the ligand environments of the magnetic site; both of which should influence the ME response with relatively small changes to the lattice constants.
Thus, they are good candidates for a theoretical study of ${\bm \alpha}$.
In this work, we present a comprehensive set of DFT calculations on these layered AFM pnictides. 
After a short review of the published information of their low temperature ground state in Sec. \ref{sec:magstruct}, we detail our DFT approach in Sec. \ref{sec:methods}.
We provide basic computed properties of the ground states in Sec. \ref{sec:basic}.
In Secs. \ref{sec:tet} and \ref{sec:trig}, we parse out the different contributions to ${\bm \alpha}$ due to spin and orbital magnetization for both the LM and CI analogs and propose some correlations.
We find that, in general, the linear ME effect in this family is driven largely by electronic contributions.
This is contrasting to the claims that the LM terms are thought to be the primary influence in the ME susceptibility.
%
%
In $\mathrm{BaMn}_2\mathrm{As}_2$, we calculate a substantial response with $\alpha_{xx} = -\alpha_{yy} \simeq 6.79$ ps/m, exceeding the response from prototypical CRO by a factor of three.
In all of our studied cases, we show that the contributions from the orbital moments cannot be neglected when calculating the purely electronic (clamped-ion) response.
This aspect is often overlooked in various studies on this phenomena.
In the trigonal materials, we find that the pnictogen substitution (from lighter to heavier elements) increases the components of ${\bm \alpha}$ alluding to the spin-orbit interaction being much more influential in this subfamily.
Additionally, we employ our described methodology on the well-studied CRO in the Appendix of this paper.
Our estimates of ${\bm \alpha}$ show good agreement with previously published values validating/benchmarking our approach.
All of our results are provided as a function of the Hubbard +U correction in the local density approximation (LDA) to DFT.
Effectively, this allows us to perform a technical probe on whether the different components of ${\bm \alpha}$ depend sensitively on the degree of electron localization or if the size of the gap at this level of theory is important for the amplitudes of the ME response.
We believe our results are useful for future experimental studies of the ME effect in this materials family and to motivate \emph{full} characterizations of ${\bm \alpha}$ in forthcoming theoretical calculations of candidate ME compounds.

\vspace*{-0.25cm}

\section{Magnetic states and symmetry\label{sec:magstruct}}

\begin{figure*}\centering 
\includegraphics[scale=0.048]{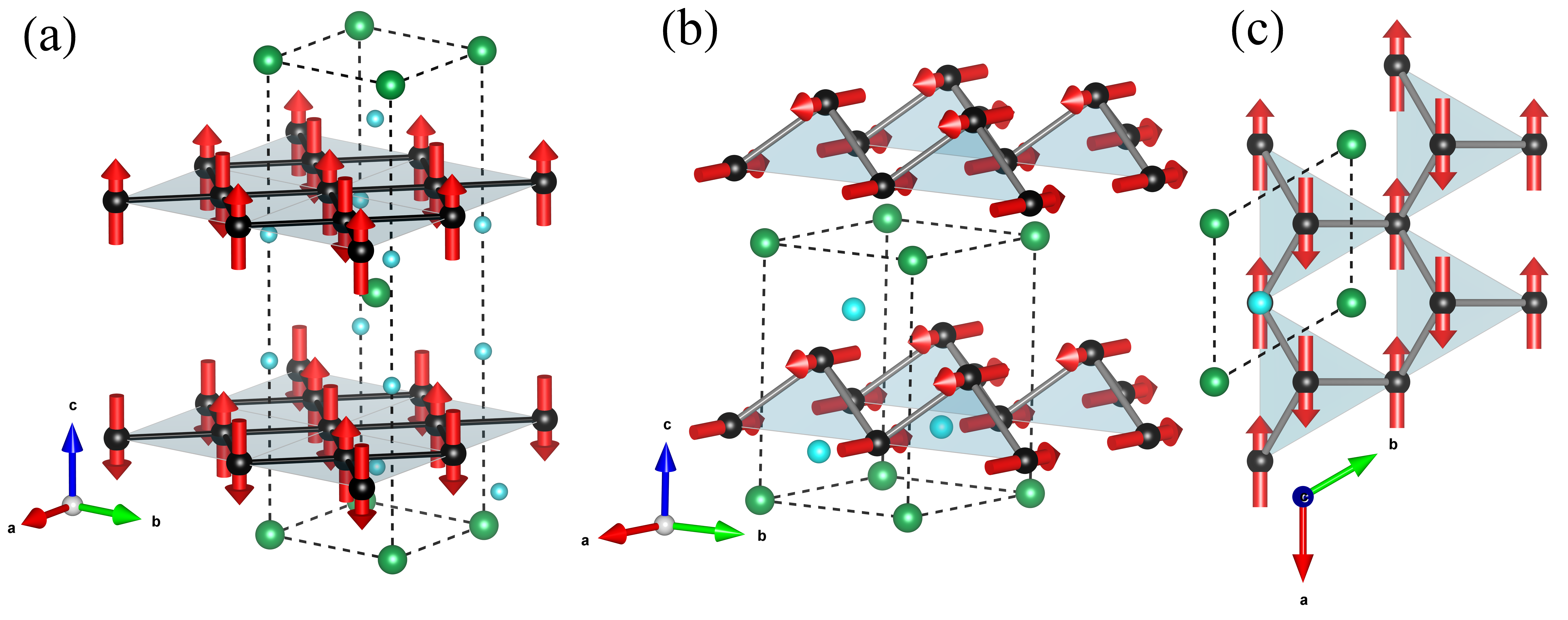} 
\caption{\label{fig:fig1} Collinear antiferromagnetic spin states in the tetragonal $I4/mmm$ (a) and trigonal $P\bar{3}m1$ (b, c) structures forming square planar and double corrugated honeycomb Mn-Mn networks respectively. The cations (Ba, Ca, Sr) are in green with the pnictogens (P, As, Sb) in cyan. The primitive cells for the tetragonal and trigonal systems are marked by a dashed black line each containing two and one formula units respectively. Additional visualization of the honeycomb lattice is provided in (c).}
\end{figure*}

We begin by discussing the ground state structure, symmetry, and magnetic states of the AFM pnictide compounds $\mathrm{(Ba,Ca,Sr)Mn}_2\mathrm{Pn}_2$ where $\mathrm{Pn} = (\mathrm{P},\mathrm{As},\mathrm{Sb})$.
Depending on the cation Ba or (Ca, Sr), the compounds crystallize at low temperature in either $I4/mmm$ or $P\bar{3}m1$ symmetry respectively \cite{Just1996}. 
The electropositive ions $\mathrm{(Ba,Ca,Sr)}$ each donate 2 electrons to the electronegative $[\mathrm{Mn}_2\mathrm{(P,As,Sb)}_2]^{-2}$ anionic heteroatomic cluster. 
This ionic bonding is a characteristic signature of the Zintl phase \cite{Bobev2006, Ovchinnikov2019} and results in a $3d^5$ ($\mathrm{Mn}^{+2}$) state \cite{Zheng1988}.
Local moment AFM behavior has been observed in all of the structures with the spin at each $\mathrm{Mn}^{+2}$ site in a range of $3$-$4$ $\mu_\mathrm{B}$/Mn which is much lower than the expected $5$ $\mu_\mathrm{B}$/Mn for the high spin $(S = 5/2)$ state\cite{Sangeetha2018,Jacobs2023}. 
Resistivity experiments have demonstrated that all of the compounds exhibit a narrow semiconducting gap where activation energies  take a range of $0.02$-$0.4$ eV \cite{Sangeetha2018, Jacobs2023}.
The onset of AFM order at N\'{e}el temperatures ($T_\mathrm{N}$) of the  $I4/mmm$ (tetragonal) compounds have been shown to depend linearly on the Pn atom substitution and the corresponding lattice constants \cite{Jacobs2023} whereas a similar correlation for the trigonal ($P\bar{3}m1$) family is unclear.
We refer to Table \ref{table:tab1} for a list of the structural parameters used in this work and reported $T_\mathrm{N}$.
The Mn-Mn network (see Fig.~\ref{fig:fig1}) differ in the two space groups. 
For the Ba-based tetragonal compounds, the magnetic sublattice is square planar with layers separated by the cations.
The $\mathrm{MnPn}_4$ tetrahedra form at edge sharing positions with a nearest-neighbor distance of approximately $3$ \AA.
In the trigonal structures, the Mn-centered tetrahedra become inverted leading to a double corrugated honeycomb layer network.
However, both the tetragonal and trigonal subfamilies are nearly isostructural themselves with pnictogen substitutions changing lattice constants by only a small amount (see Table \ref{table:tab1}).
For the tetragonal compounds, the minimum energy configuration is a layered AFM G-type checkerboard pattern with spins arranged normal to the square magnetic plane along the tetragonal ($\mathbf{c}$) axis of the primitive cell - see Fig.~\ref{fig:fig1} (a).
Therefore, the magnetic order has both inter- and intralayer AFM couplings. 
By symmetry of the magnetic space group ($I4'/m'm'm$, No. 139.536), the linear ME tensor must take the form\cite{Watanabe2018}
\begin{align}\label{eq:I4pmpmpm}
{\bm \alpha} (I4'/m'm'm)= \begin{pmatrix}
\alpha_{xx} & 0 & 0\\
  &  & \\
0  & -\alpha_{xx} & 0\\
  &  & \\
0 & 0 & 0 \\
\end{pmatrix}
\end{align}
with only one independent coefficient.
Here, it is evident that the ME property is two-dimensional (2D) - admitting only a \emph{transverse} response in the basal plane of the crystal.
%
%

In the closely-related trigonal $(P\bar{3}m1)$ materials, the spins N\'{e}el order along the $\mathbf{a}$ axis.
The spin order on the triangular lattice is intralayer aligned (ferromagnetic coupling) with nearest-neighbor layers anti-aligned giving AFM coupling.
See Fig. 1 (b) and (c) for a visual representation.
Applying the group operations of the magnetic space group\cite{nonstandNote} ($C2'/m$, No. 12.60) yields the following four independent (nonzero/allowed) coefficients of the linear ME tensor,
%
%
\begin{align}\label{eq:C2pm}
{\bm \alpha} (C2'/m) = \begin{pmatrix}
0 & \alpha_{yx} & \alpha_{zx}\\
  &  & \\
\alpha_{xy} & 0 & 0\\
  &  & \\
\alpha_{xz} & 0 & 0 \\
\end{pmatrix}.
\end{align}
%
%
%
It should be noted that the relative signs of the antisymmetric off-diagonal ${\bm \alpha}$ are not defined within the symmetry consideration and instead depend on the detailed microscopic information which we will uncover later in this paper.
From looking at Eq.~(\ref{eq:C2pm}) with respect to the spin axis direction $\mathbf{a} || \mathbf{x}$, we have two coefficients ($\alpha_{yx}$, $\alpha_{zx}$) that correspond to a transverse response whereas the other two components ($\alpha_{xy}$, $\alpha_{xz}$) are longitudinal.
%
%
We list the forms of Eq.~(\ref{eq:I4pmpmpm}) and (\ref{eq:C2pm}) in Table \ref{table:tab1} for reference.

\begin{table}[tb]
\begin{ruledtabular}
\begin{tabular}{c c c @{\hspace{-0.01cm}}c @{\hspace{-0.825cm}}c @{\hspace{-0.75cm}}c c c }
  & a = b  & c& $\hat{\bm \alpha}$  &  \hspace*{20pt}\underline{Wyckoff $(z_0)$} & &  $T_\mathrm{N}$ & \\ 
 & & & & & & & \\
  $I4/mmm$& & & &   Mn &   Pn & &\\
 & & & & & & & \\
$\mathrm{BaMn}_2\mathrm{P}_2$  & 4.037&  13.052& &  1/4 & 0.3548 & 795 &  \\ 
$\mathrm{BaMn}_2\mathrm{As}_2$   & 4.154 & 13.449 & $\begin{pmatrix}
    \filledDot & \smallDot & \smallDot \\
    \smallDot & \emptyDot & \smallDot \\
    \smallDot & \smallDot & \smallDot \\
\end{pmatrix}$& 1/4 & 0.3613 & 618&  \\ 
$\mathrm{BaMn}_2\mathrm{Sb}_2$   & 4.397 & 14.330 & &  1/4 & 0.3659 & 450 &  \\ 
& & & & & & & \\
  $P\bar{3}m1$& & & & Mn& Pn &\\
  & & & & & & & \\
$\mathrm{CaMn}_2\mathrm{P}_2$   & 4.096&  6.848& & 0.6246 & 0.2612    & 70 &  \\ 
& & & & & & & \\
$\mathrm{CaMn}_2\mathrm{As}_2$  & 4.230&  7.033& & 0.6237&  0.2557   & 62 &  \\ 
$\mathrm{CaMn}_2\mathrm{Sb}_2$  & 4.525&  7.443& $\begin{pmatrix}
    \smallDot & \filledDot & \filledDot \\
    \filledDot & \smallDot & \smallDot \\
    \filledDot & \smallDot & \smallDot \\
\end{pmatrix}$& 0.6221&   0.2490   & 85 & \\ 
$\mathrm{SrMn}_2\mathrm{P}_2$  & 4.156&  7.096&  & 0.6005&   0.2496  & 53&  \\ 
& & & & & & &\\
$\mathrm{SrMn}_2\mathrm{As}_2$  & 4.296&  7.30& & 0.6231 &  0.2667   & 120 & \\ 
& & & & & & &\\
$\mathrm{SrMn}_2\mathrm{Sb}_2$ & 4.580 & 7.73& & 0.6194  &  0.2625   & 110 &  \\ 
\end{tabular}
\end{ruledtabular}
\caption{\label{table:tab1}%
Measured structural parameters used in this work. Lattice constants are given in \AA. Wyckoff positions of cations (Ba, Ca, Sr) are situated at (0,0,0) whereas the $\{\mathrm{Mn}, \mathrm{Pn}\}$ atoms are located at $\{(0, 1/2, z_0),(0, 0, z_0)\}$ and $\{(1/3, 2/3, z_0), (1/3, 2/3, z_0)\}$ for the tetragonal and trigonal settings respectively. The N\'eel temperatures $T_\mathrm{N}$ are given in K. We also provide the symmetrized $\hat{ \bm\alpha}$ tensor with independent and dependent (negative) coefficients marked by ($\filledDot$\,) and ($\emptyDot$\,) respectively. Properties are compiled from a number of references\cite{Bobev2006, Sangeetha2016, Sangeetha2018, Sangeetha2021, Jacobs2023}.
}
\end{table}

With $\mathrm{Mn}^{+2}$ formally being $3d^5$ occupancy, Hund's rule tells us that we should expect negligible or quenched orbital magnetization.
%
%
However, due to the ligand $\mathrm{MnPn}_4$ environments, hybridization between the Mn $d$-band with the pnictogen $p$-valence occurs in $\mathrm{BaMn}_2\mathrm{(As,Sb)}_2$ as verified in a angle-resolved photoemission spectroscopy (ARPES) experiments \cite{Zhang2016} and from DFT\cite{An2009}.
It has been proposed that $p$-$d$ hybridization can lead to ME effects \cite{Murakawa2010, Lovesey2018}.
To what extent that this plays a role in the ME response for the layered pnictides is one aspect of this current investigation.
This materials family also includes the pnictide Bi with interesting magnetoresistance properties\cite{Huynh2019}.
However, since Bi is a much heavier atom with a stronger spin-orbit interaction than the other pnictogens, we had difficulty producing a gapped state with our DFT prescription.
Therefore, we have decided to neglect the Bi-based compounds in our analysis.
%
%
%
We should also note that some experimental evidence suggests that some of the pnictides, such as $\mathrm{CaMn}_2\mathrm{P}_2$\cite{Islam2023}, $\mathrm{CaMn}_2\mathrm{As}_2$\cite{Ding2021}, or $\mathrm{CaMn}_2\mathrm{Sb}_2$\cite{Bridges2009} may have broken collinear or incommensurate order. 
However, we make the assumption that all materials investigated in this work have collinear AFM ground states as that this should provide a reasonable approximation to the computed values of ${\bm \alpha}$. 
In the next sections, we detail our DFT methodology to fully characterize the linear ME response.
%


\section{Methods\label{sec:methods}}

\subsection{Density functional theory specifics\label{sec:methodsDFT}}

We implement our density functional theory (DFT) calculations within the local density approximation (LDA) using the open-source electronic structure package GPAW \cite{Mortensen2024} and the Atomic Simulation Environment (ASE) \cite{Larsen2017}.
All calculations (at T$ = 0$ K) utilize a plane-wave basis set with an energy cut-off of 800 eV and a $\Gamma$-centered $k$-space grid of 10$\times$10$\times$10 points.
We consider one and two formula unit cells for the $P\bar{3}m1$ and $I4/mmm$ structures respectively.
%
%
The single-particle wavefunctions are approximated within the projector-augmented wave (PAW) method \cite{Blochl1994} solving explicitly for the 10 valence $e^-$ of cations Ba ($5s^25p^66s^2$), Sr ($4s^24p^65s^2$), and Ca ($3s^23p^64s^2$), and the 5 valence $e^-$ for P ($3s^23p^3$) and As ($4s^24p^3$).
Ten additional semi-core $e^-$ are included in the case of Sb ($4d^{10}5s^25p^3$). 
The Mn atom is treated with 15 valence electrons ($3s^23p^63d^54s^2$).
Cut-off radii $(r_\mathrm{cut})$ of the PAW spheres for the different species are set acccording to GPAW defaults \cite{noteRcut}.
We used a Fermi-Dirac distribution with a smearing of 1 meV for the electronic occupation numbers. %
To better describe the electronic structure, the rotationally invariant Hubbard +U correction \cite{Dudarev1998} is applied to the $3d$ states of the Mn atom and several of the results below will be shown as a function of U ranging between 0 and 3 eV.

\subsection{Lattice-mediated spin}\label{sec:methodsLMS}

We first perform a spin-polarized relaxation of the internal coordinates with lattice constants fixed to those measured from experiment (see Table \ref{table:tab1}).
This allows us to obtain a force-free reference configuration of the AFM ground state converged to a maximum force tolerance on the atoms of 0.001 eV/\AA.
Expanding the energy of the crystal in the harmonic approximation, one can show that the atomic displacements corresponding to those from an external $\mathbfcal{E}$ that is \emph{static} and homogeneous are\cite{Wu2005, Iniguez2008, Ye2014} 
%
%
\begin{align}\label{eq:frozenE}
u_\beta(\mathbfcal{E}) = \Omega_0^{-1}\sum_{j\kappa}\left(K_{\beta\kappa}\right)^{-1} Z_{j\kappa}^e \mathcal{E}_j
\end{align}
where $\mathbf{K}$ is the force constant matrix, $\Omega_0$ the cell volume, and $Z_{j\kappa}^e = \Omega_0 \,\partial P_j / \partial u_\kappa$ are the Born effective charges (BECs). 
Greek symbols run over atomic coordinates ($\beta = 1,...,3N_a$) and Latin indices define Cartesian reference frame $j = x,y,z$. 
With linear response methods\cite{Baroni2001}, we compute the BECs and also the $\Gamma$-point phonon spectra to obtain the force constant matrix.
The pseudoinverse of $\mathbf{K}$ is computed by the Moore-Penrose technique \cite{Strang80} which effectively traces out the acoustic phonon eigenmodes that make it singular \cite{Wu2005}.
We should mention that we find negligible influence of computing converged BECs within collinear spin-polarized or noncollinear (with SOC) approaches using an atomic position shift of $0.01$ \AA. 
Therefore, we present results using the spin-polarized mode. 
%
%

%
Structures were displaced under $\mathbfcal{E}$ in all three Cartesian unit directions $\mathbf{x},\mathbf{y},\mathbf{z}$ (where i.e. $\mathbf{x}||\mathbf{a}$, $\mathbf{y}||\mathbf{b}$, and $\mathbf{z}||\mathbf{c}$ in the special case of the $I4/mmm$ symmetry).
The amplitude of the field $|\mathbfcal{E}|$ was considered to be less than $0.01$ V/nm to ensure predictions are in the linear limit.
Spin orbit coupling (SOC) was included and the self-consistent field (SCF) loop was terminated after a maximum absolute change in the integrated electronic density is less than $10^{-9}$ electrons/valence $\mathrm{e}^{-}$.
The quantity $\mathbf{M}^\mathrm{S}$, denoted as the spin magnetization, is computed from an integral of the spin density over the unit cell and divided by its volume.
The resulting linear change in the total spin magnetization of the cell with respect to $\mathbfcal{E}$ was fitted giving ${\bm \alpha}$ in appropriate SI units (ps/m). 
Our convergence criteria yielded a resolution of the ME tensor components (and symmetry) to within $\pm 10^{-3}$ ps/m.
In Sec.~\ref{sec:results}, we will denote the lattice-mediated (LM) spin contribution by:
$$\alpha_{ij}^{\mathrm{LM},S} \equiv \mu_0 \left(\frac{\partial M_i^\mathrm{S}}{\partial \mathcal{E}_j}\right)\bigg|_{\mathbfcal{B}=0}.$$
We find that under the time reversal operation, $\mathbf{m}_a^\mathrm{S} \to -\mathbf{m}_a^\mathrm{S}$.
As this indicates a different AFM domain orientation, the signs of all components of ${\bm \alpha}^\mathrm{LM,S}$ also change sign as discussed in Ref. [\citen{Bousquet2024}] in the case of CRO.
Therefore, we only present ${\bm \alpha}$ for a single AFM domain.
In all of the simulations, we arrange this state consistently such that all systems have the same spin orientation on their symmetric Wyckoff positions.
Finally, it is common to also see values presented in the literature in Gaussian units, to which the conversion is $1$ ps/m equal to $\sim 3\times 10^{-4}$ g.u.
See Ref. [\citen{Rivera1994}] for a review of different units used in ME measurements.
\subsection{Lattice-mediated orbital magnetization}\label{sec:methodsLML}

Next, we turn to the LM orbital moment contribution to the ME tensor.
By analyzing the Berry phases, the orbital magnetization can be evaluated at the level of the entire unit cell using the modern theory of orbital magnetization\cite{Thonhauser2005, Resta2010}.
An alternative approach, which is employed here, is to evaluate quantities locally involving the dimensionless angular momentum operator $\hat{\mathbf{L}} = \hat{\mathbf{r}}\times\hat{\mathbf{p}}/\hbar$.
We should note that operators of the form $\hat{\mathbf{r}}$ are ill-defined in the periodic unit cell but not when restricted to the PAW spheres.
Since the all-electron partial waves are exact numerical quantities within the PAW spheres, then physical observables that are quite localized (i.e. $\mathbf{m}^\mathrm{S}_a$ and its orbital counterpart) can be extracted accurately.
The PAW approach to compute orbital moments has been shown to agree quite well with predictions from the modern theory for a variety of materials \cite{Mortensen2024}.
%
%

%
To compute the orbital magnetic moments, we use the following expression\cite{Mortensen2024},
\begin{align}\label{eq:orb}
    \mathbf{m}_{a}^\mathrm{L} = \frac{\mu_\mathrm{B}}{N_k} \sum_{\mathbf{k}n} f_{\mathbf{k}n} \left\langle \psi_{\mathbf{k}n}^a\right| \hat{\mathbf{L}}\left| \psi_{\mathbf{k}n}^a \right\rangle,
\end{align}
with $\mu_\mathrm{B}$ the Bohr magneton, $N_k$ the number of $\mathbf{k}$-points, and $f_{\mathbf{k} n}$ the band occupancy.
The all-electron wavefunctions $|\psi_{n}^a\rangle$ are defined within the atom-centered PAW sphere (index $a$) in the usual way \cite{Mortensen2024}.
%
%

%
Using the LM approach outlined in the previous section, the orbital magnetization per unit cell $(\mathbf{M}^\mathrm{L} = \Omega_0^{-1} \sum_a \mathbf{m}_{a}^\mathrm{L})$ is computed from Eq. (\ref{eq:orb}) as a function of the frozen displacements due to $\mathbfcal{E}$.
%
%
Similarly to the previous section, we indicate the LM orbital component as, 
\begin{align}\label{eq:orb_contr}
\alpha_{ij}^{\mathrm{LM},L} \equiv \mu_0 \left(\frac{\partial M_i^{L}}{\partial \mathcal{E}_j}\right)\bigg|_{\mathbfcal{B}=0}
\end{align}
which may also be separated into local $d-$ or $p-$orbital contributions. 
\subsection{Clamped-ion spin-driven polarization}\label{sec:methodsCIS}

To compute the clamped-ion (CI) component of the ME tensor, we utilize an additional term in the Hamiltonian that couples directly to the spin operator $\mathbf{S}$, $H = -\mathbfcal{B} \cdot \mathbf{S}$.
Naturally, this requires the SOC to be included in the SCF cycle of the DFT calculation.
The ionic positions are fixed and the SCF cycle is converged to within a maximum absolute change in the electronic density ($10^{-9}$ electrons/valence $\mathrm{e}^{-}$) under a finite $\mathbfcal{B}$.
We calculate the total polarization $\mathbf{P}$ of the computational cell utilizing the Berry phase approach \cite{King-Smith1993, Resta1994}.
The external magnetic field in all three Cartesian unit directions $(\mathbf{x},\mathbf{y},\mathbf{z})$ is swept from $\pm 1$ T in finite steps and $\mathbf{P}$ is recorded at each field magnitude.
%
%
%
We verify that in all cases presented the systems remain gapped which is a requirement in the Berry phase method to compute the electronic polarization. 
Our notation for this term is,
\begin{align}\label{eq:CI_S}
\alpha_{ij}^{\mathrm{CI},S} \equiv \mu_0 \left(\frac{\partial P_j}{\partial \mathcal{B}_i}\right)\bigg|_{\mathbfcal{E}=0},
\end{align}
with the derivative computed from finite differences.

\subsection{Clamped-ion orbital magnetization}\label{sec:methodsCIL}

Analogous to the CI spin component, there is expected to be a ME response due to the relative motion of orbital currents under $\mathbfcal{E}$\cite{Malashevich2012}.
To compute this, we implement an approximation.
We perform a first-order variation of Eq.~(\ref{eq:orb}) with respect to a perturbation potential of the form $V = e \mathbf{r}\cdot\mathbfcal{E}$ with $e > 0$ to obtain the contribution to ${\bm \alpha}$ due to clamped-ion orbital moments,
\begin{widetext}
\begin{align}\label{eq:orb_CIL}
    \alpha_{ij}^\mathrm{CI, L} &\equiv \frac{\mu_0}{\Omega_0} \sum\limits_a \left(\frac{\partial m^\mathrm{L}_{a,i}}{\partial \mathcal{E}_j}\right)\bigg|_{\mathbfcal{B}=0} =\frac{2i \mu_0 \mu_\mathrm{B}^2}{N_k \Omega_0} \sum\limits_a \sum_{n\neq m} \sum_{\mathbf{k}} (f_{\mathbf{k}n}-f_{\mathbf{k}m}) \frac{\langle \psi_{\mathbf{k} n}^a|\hat{L}_j| \psi_{\mathbf{k} m}^a\rangle\langle \psi_{\mathbf{k} m}|\hat{p}_i|\psi_{\mathbf{k} n}\rangle}{(\varepsilon_{\mathbf{k}n}-\varepsilon_{\mathbf{k}m})^2}.
\end{align}
\end{widetext}
where $L_r$ and $p_s$ are the usual angular and linear momentum operators in the Cartesian representation $i,j = x,y,z$.
In this form, Eq.~(\ref{eq:orb_CIL}) is given in inverse velocity with $\langle {\bm \hat{L}} \rangle$ dimensionless.
%
%
The prefactor contains $i/N_k$ the imaginary number divided by number of $\mathbf{k}$-points used.
The summation $n \neq m$ is over all occupied and \emph{unoccupied} bands with their associated energy eigenvalues $\varepsilon_{\mathbf{k}m}$ and all-electron spinor wavefunctions $\psi_{\mathbf{k}m}$. 
The term $\alpha_{ij}^\mathrm{CI, L}$ is computed from the zero-field ground states and largely corresponds to the "cross-gap" contribution derived in Ref. [\citen{Essin2010}].
We should comment on how this term compares to results derived from the modern theory of orbital magnetization presented in Ref.~[\citen{Malashevich2012}].
In that work, the ME response is separated into local and itinerant circulation parts in addition to an isotropic Chern-Simons term.
We expect that Eq.~(\ref{eq:orb_CIL}) should give similar results to the sum of the local and itinerant circulations but we do not include the Chern-Simons part which is expected to be small.
Also, since  Eq.~(\ref{eq:orb_CIL}) only integrates the orbital angular momentum over the PAW spheres, we do not consider interstitial regions which may contribute to the orbital CI ME response (as well as the orbital LM part discussed earlier).

\begin{figure*}\centering
\hspace*{-2.5pt}\includegraphics[scale=0.235]{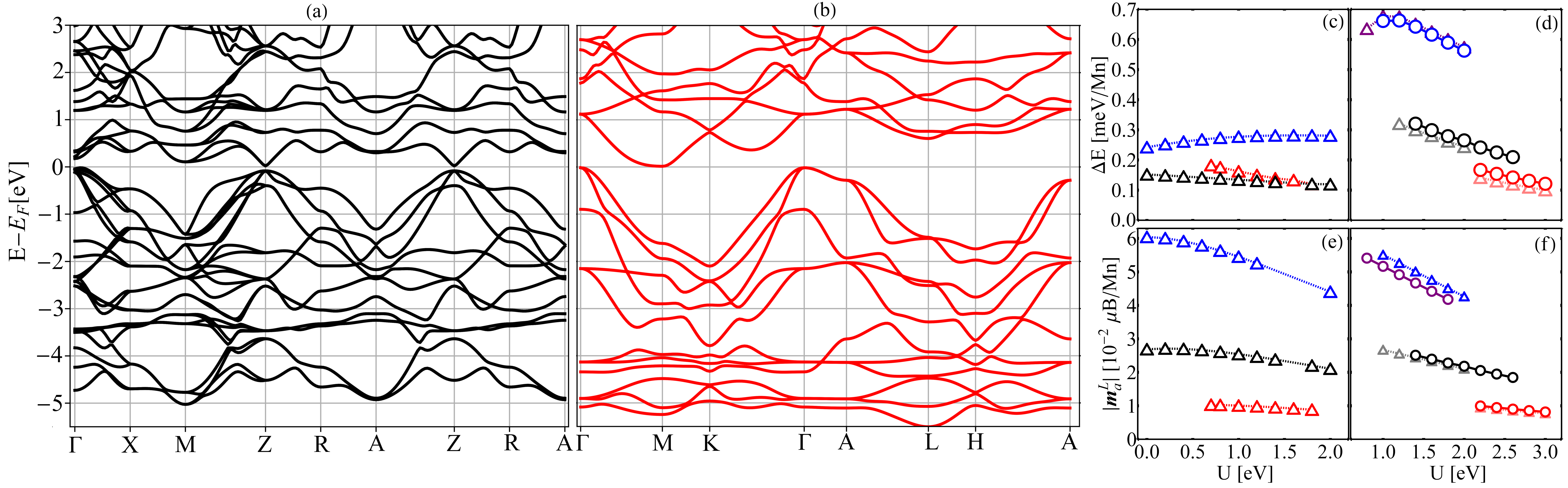}%
\caption{\label{fig:basic} Representative electronic band structure across the Brillouin zone in $\mathrm{BaMn}_2\mathrm{As}_2$ with bare LDA in (a) and $\mathrm{CaMn}_2\mathrm{P}_2$ with $U = 2.4$ eV in (b). 
The magnetic anisotropy $\Delta E$ for $I4/mmm$ and $P\bar{3}m1$ compounds are shown in the panels of (c) and (d) accordingly given in meV per Mn site as a function of $U$. The bottom panels display the ground state $|\mathbf{m}_a^\mathrm{L}|$ on a single Mn site for tetragonal (e) and trigonal materials (f) respectively.}
\end{figure*}

\subsection{Error metrics}

We can discuss some quantifications of error/precision that may arise in our calculations.
First, is the expectation of symmetry in the raw data. 
In the case of the Ba-based compounds, we expect that $\alpha_{xx} = - \alpha_{yy}$ with all other components zero which is evident in our results within $\pm 10^{-6}$ ps/m precision.
For the trigonal systems, the precision is relaxed slightly, but we still obtain the expected symmetry of Eq.~(\ref{eq:C2pm}) within $\pm 10^{-3}$ ps/m.
The second is comparison to another method.
Using the converse Zeeman field effect the total spin component of the ME response, ${\bm \alpha}^\mathrm{S} =  {\bm \alpha}^\mathrm{LM, S} + {\bm \alpha}^\mathrm{CI, S}$, can be inferred by relaxing the ionic positions in an external field $\mathbfcal{B}$ with self-consistent SOC calculations.
This has been demonstrated in other publications for CRO \cite{Bousquet2011, Malashevich2012, Bousquet2024}.
We performed converse Zeeman field calculations for a few select +U values (both for the data presented in the Appendix and in the main text) showing good agreement in our computed ${\bm \alpha}^\mathrm{S}$ thus validating our method for the spin components.
%
%

\section{Results}\label{sec:results}
%


\subsection{Basic properties: band structure, anisotropy, and orbital moments\label{sec:basic}}

In the following section, we use red, black, and blue for data points regarding pnictogen content $\mathrm{P}, \mathrm{As}$, and $\mathrm{Sb}$ respectively.
Each data point represents an independent DFT relaxation and corresponding workflow set out in Sec. \ref{sec:methods}.
With the exception of $\mathrm{BaMn}_2\mathrm{(As,Sb)}_2$, the bare LDA calculations predict metallic states for all AFM configurations studied. 
The method of the Hubbard correction opens a gap in all of the trigonal materials and in the case of $\mathrm{BaMn}_2\mathrm{P}_2$.
Representative spin-polarized electronic band diagrams of the AFM ground states of $\mathrm{BaMn}_2\mathrm{As}_2$ (bare LDA) and $\mathrm{CaMn}_2\mathrm{P}_2$ (U = 2.4 eV) are displayed in Fig.~\ref{fig:basic} (a) and (b) respectively.
In the $I4/mmm$ compounds, the gap ($E_g$) is direct located at $Z (\frac{1}{2},\frac{1}{2},\frac{1}{2})$. 
For the band structure shown in Fig.~\ref{fig:basic} (a), the band gap is $40$ meV.
The topology is slightly different in the trigonal structures with an indirect 68 meV gap opening in the case of (b).
The conduction band minimum is located at M $(\frac{1}{2}, 0, 0)$ whereas the valence band maximum appears at $\Gamma$. 
Comparable diagrams are obtained by pnictide (or $\mathrm{Ca}\to\mathrm{Sr}$) substitution. 
Both (a) and (b) highlight the narrow gap and semiconducting nature of these compounds.
Throughout Sec. \ref{sec:results}, we will present our results as a function of the Hubbard +U.
But before doing so, we should comment on which of our calculations corresponds best to the true ground state. 
For this, we turn to the computed $E_g$ as a guide.
For the Ba-based materials, we find $E_g = (0.03, 0.05, 0.15)$ eV with $U = (0.7, 0.0, 0.8)$ eV for Pn = (P, As, Sb) respectively.
These values of the band gap are in reasonable agreement with experimental evidence\cite{Bobev2006, Sangeetha2016, Sangeetha2021, Jacobs2023}.
The zero-field spin magnetic moments on each Mn site for the best estimate ground states - $m^\mathrm{S}$ = (3.27, 3.24, 3.82) $\mu_\mathrm{B}$/Mn - also agree with the observed departure from the expectation in $S = 5/2$ magnetism (see Refs.~[\citen{An2009, Sangeetha2018}]). We note however, that the local moments are somewhat ambiguous in experiments as well as theory and the comparison should thus merely be regarded as a rough sanity check.
%
%
The same analysis can be made for the trigonal materials.
We list our best estimates of their spin moments and gaps in Table \ref{table:tab2} along with the +U used.
In Fig.~\ref{fig:basic} (c, d) (top panels), we provide estimates of the anisotropy energy $\Delta E$ by comparing the differences of the total energy $E$ from DFT calculation with AFM spin orientation aligned along different direction.
We use $\Delta E = E(\pm\mathbf{m}_a||\mathbf{c})-  E(\pm\mathbf{m}_a||\mathbf{a})$ and $\Delta E = E(\pm\mathbf{m}_a||\mathbf{a})-  E(\pm\mathbf{m}_a||\mathbf{c})$ for both the tetragonal and trigonal systems correspondingly as a function of the onsite Hubbard correction.
All states are assumed to have an AFM configuration (hence $\pm$ in our notation).
One can appreciate from this data that the easy-axis is always out-of-plane for the Ba-based compounds whereas the minimum energy is always for spin along the $\mathbf{a}$ axis in the trigonal structures.
%
%
The trend of $\Delta E$ with increasing +U is to gradually reduce in all of the compounds (with the exception of $\mathrm{BaMn}_2\mathrm{Sb}_2$).
When $\mathrm{Sr}$ is substituted into the calcium site, the anisotropy energy is slightly reduced in all cases with a similar trend (see pink, gray, and purple data points for respective P, As, and Sb pnictogens).
%

%

%
Also provided in Fig.~\ref{fig:basic} (e,f) (bottom panels) is our estimates of $|\mathbf{m}_a^\mathrm{L}|$ at a single Mn site for all compounds using Eq.~(\ref{eq:orb}).
The direction of the \emph{net} orbital Mn (site) moments in the calculations are found to be parallel to that of the Mn spin.
The value is quite low, i.e. in $\mathrm{BaMn}_2\mathrm{Sb}_2$, of $|\mathbf{m}_a^\mathrm{L}| = 0.056$ $\mu_\mathrm{B}$/Mn demonstrating the quasi-quenched orbital nature due to the half-filled $d$-orbitals of manganese.
By separating out the $p$- and $d$-bands in Eq.~\ref{eq:orb}, we find that in this case, approximately 0.067 $\mu_\mathrm{B}$/Mn comes from the $d$-band whereis a negative contribution of around $-0.011$ $\mu_\mathrm{B}$/Mn is due to $p$-character (around 17$\%$).
Formally, lone $\mathrm{Mn}^{+2}$ should not have occupancy of the $4p$ state, but it is obtained due to band overlap with the neighboring pnictogen ligands.
This contribution is much larger in the P and As systems (best estimates), as one finds about $43\%$ and $25\%$ of the orbital magnetization respectively coming from the $4p$ electrons.

In the case of the trigonal structures, the net site orbital moment is also very small. 
For example, in $\mathrm{CaMn}_2\mathrm{As}_2$ at $U = 1.6$ eV, we find $0.032$ $\mu_\mathrm{B}$/Mn from $d$-orbitals and $-0.008$ $\mu_\mathrm{B}$ from the $p$-orbitals.
Upon pnictogen substitution, the relative $p$ contribution is approximately the same which is in contrast to the Ba-based materials.
Similar trends are obtained for the strontium substitution (see pink, gray, and purple data points).
The influence of the Hubbard +U causes $|\mathbf{m}_a^\mathrm{L}|$ to decrease in all structures. 
%
%
%

\begin{table}[b!]
\begin{ruledtabular}
\begin{tabular}{@{\hspace{-0.03cm}}c @{\hspace{-0.9cm}}c  @{\hspace{-2.4cm}}c  @{\hspace{-1.65cm}}c @{\hspace{-0.05cm}}c c c c c}
 & \hspace*{20pt}\underline{\,\,\,\,\,\,\,\,\,\,Mn\,\,\,\,\,\,\,\,\,\,} &  \hspace*{72pt}\underline{\,\,\,\,\,\,\,\,\,\,Pn\,\,\,\,\,\,\,\,\,\,} &  & &   &  & \\ 
& & & & & & & \\
& $m^\mathrm{S}$ & $m^\mathrm{L}$ &  $m^\mathrm{S}$ & $m^\mathrm{L}$    & $\Delta E$  &  $E_g(\Delta)$ & +U\\ 
& & & & & & &\\
$\mathrm{BaMn}_2\mathrm{P}_2$  & 3.274 & 0.010 & 0 & 0 & 0.18 & 31(24)& 70\\
& & & & & & &\\
$\mathrm{BaMn}_2\mathrm{As}_2$ & 3.235 & 0.027 & 0 & 0 & 0.15 & 40(27) & 0\\ 
& & & & & & &\\
$\mathrm{BaMn}_2\mathrm{Sb}_2$ & 3.815 & 0.056 &  0 &  0 & 0.27 & 150(160) & 80\\ 
& & & & & & &\\
$\mathrm{CaMn}_2\mathrm{P}_2$  & 4.129 & 0.009 & -0.006 & 0.001 & 0.15 &  68 (88) & 240\\
& & & & & & &\\
$\mathrm{CaMn}_2\mathrm{As}_2$  & 4.077 & 0.024 & -0.005 & 0.002 & 0.30&  87(62)& 160\\ 
& & & & & & &\\
$\mathrm{CaMn}_2\mathrm{Sb}_2$  & 4.171 & 0.048 & -0.002 & 0.004 & 0.62 & 93(70)& 160\\ 
& & & & & & &\\
$\mathrm{SrMn}_2\mathrm{P}_2$  & 4.187& 0.009 & -0.002 & 0.001 & 0.12 & 118(124)& 260\\ 
& & & & & & &\\
$\mathrm{SrMn}_2\mathrm{As}_2$ & 4.099 & 0.023 & -0.001 & 0.002 & 0.28 & 95(85)& 160\\ 
& & & & & & &\\
$\mathrm{SrMn}_2\mathrm{Sb}_2$ & 4.483 & 0.024 & 0.008  & 0.002 & $0.39$ & $318$(350)& 380\\  
& & & & & & &\\
\end{tabular}
\end{ruledtabular}
\caption{\label{table:tab2}%
Ground state spin and orbital moment components along the easy-axis for both Mn and Pn Wyckoff locations listed in Table~\ref{table:tab1} (in $\mu_\mathrm{B}$ per site). In units of meV, we provide the magnetic anisotropy energy $\Delta E$ (per Mn atom), the band gap $E_g$ (with SOC), the chosen Hubbard +U correction from our best estimate calculation, and measured activation energies ($\Delta$) obtained from experiment\cite{Bobev2006, Sangeetha2016, Sangeetha2018, Sangeetha2021, Jacobs2023}.}
\end{table}








%

%
Before the computed ME properties of the family are discussed, some quantities are useful for analyzing the data.
We sum the LM and CI parts individually as,
\begin{align}\label{eq:LMCI}
    {\bm \alpha}^\mathrm{LM} &= {\bm \alpha}^\mathrm{LM,S} + {\bm \alpha}^\mathrm{LM,L} \\ \nonumber
    {\bm \alpha}^\mathrm{CI} &= {\bm \alpha}^\mathrm{CI,S} + {\bm \alpha}^\mathrm{CI,L} \\ \nonumber
\end{align}
and the spin (S) and orbital components (L) with
\begin{align}\label{eq:SL}
    {\bm \alpha}^\mathrm{S} &= {\bm \alpha}^\mathrm{LM,S} + {\bm \alpha}^\mathrm{CI,S} \\ \nonumber
    {\bm \alpha}^\mathrm{L} &= {\bm \alpha}^\mathrm{LM,L} + {\bm \alpha}^\mathrm{CI,L}. \\ \nonumber
\end{align}
This allows us to identify a magnetoelectric participation ratio (MEPR) of the various processes proportioning a given component to that of the total amplitude which is the sum of all contributions
\begin{align}\label{eq:total}
    {\bm \alpha} &= {\bm \alpha}^\mathrm{LM,S} + {\bm \alpha}^\mathrm{LM,L} + {\bm \alpha}^\mathrm{CI,S} + {\bm \alpha}^\mathrm{CI,L}.\\ \nonumber
\end{align}

\begin{figure}\centering
\includegraphics[scale=0.36]{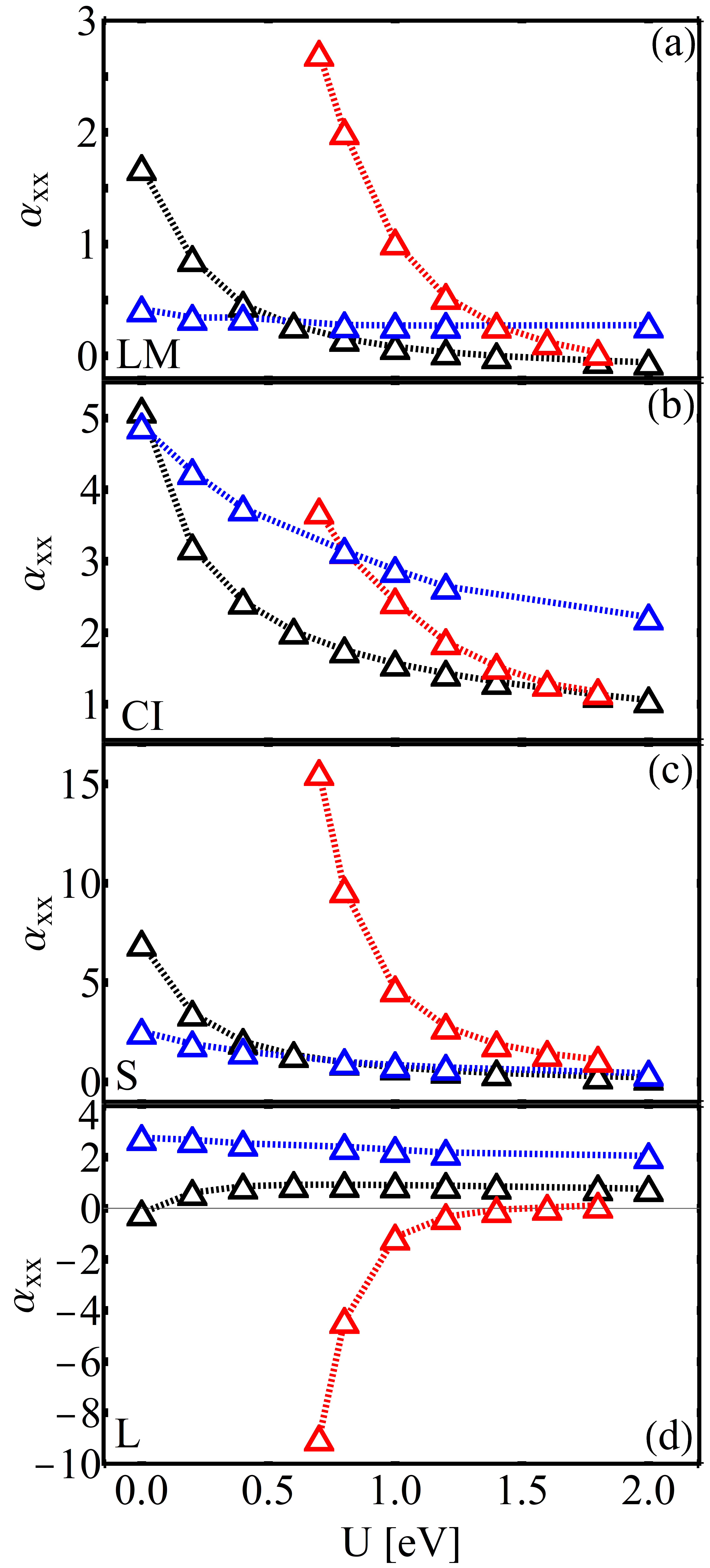}
\caption{\label{fig:fig_tet_LM_CI_L_S} Calculated transverse ME response in the $I4/mmm$ structures for P, As, Sb (red, black, blue) respectively. The decomposed contributions in panel (a), (b), (c) and (d) correspond to LM, CI, S and L modes accordingly. Units of $\alpha_{xx} = - \alpha_{yy}$ are given in ps/m.}
\end{figure}

\subsection{Magnetoelectricity in the tetragonal compounds\label{sec:tet}}

According to $I4'/m'm'm$ symmetry, it is expected that only one independent coefficient of ${\bm \alpha}$ is nonzero.
Within precision, we find all calculated contributions as well as the sum of the parts obey $\alpha_{xx} = - \alpha_{yy} \neq 0$ (along with expected zero components). 
We verify numerically that under time-reversal, the spin magnetization flips $180^\circ$ (to a different AFM domain orientation) subsequently changing the sign of ${\bm \alpha}$ (and all its separate contributions).
In Fig.~\ref{fig:fig_tet_LM_CI_L_S}, we use Eq.~(\ref{eq:LMCI}) and (\ref{eq:SL}) to decompose $\alpha_{xx}$ as a function of the Hubbard correction.
We plot separately the LM, CI, spin and orbital contributions in (a), (b), (c) and (d) respectively.
One can appreciate that the various components of ${\bm \alpha}$ are highly dependent on the Hubbard U.
For example, the clamped-ion spin response of $\mathrm{BaMn}_2\mathrm{P}_2$ decreases by nearly $67\%$ if +U is changed from 0.7 to 0.8 eV.
In all of the $I4/mmm$ compounds, we find that the largest (positive sign) involvement in the ME response is due to the ${\bm \alpha}^\mathrm{CI, S}$ term - see panels (b) and (c). 
However, we see a large contribution also from ${\bm \alpha}^\mathrm{CI, L}$ albeit negative sign in the case of $\mathrm{BaMn}_2\mathrm{P}_2$
%

%
These best estimates of $\alpha_{xx}$ are chosen at a given +U by examining the band gaps of the ground state as discussed in the previous section.
In this fashion, these values may be closest to the ones observed experimentally at T$ = 0$ K in a single-crystal in the absence of any domain state inhomogeneities.
We list these values in Table~\ref{table:tab3}. 
Hence using the computed band gap as a guide, we expect $\mathrm{BaMn}_2\mathrm{As}_2$ to have the largest ME response ($\alpha_{xx} \approx 6.79$ ps/m) in this family.
In (a), (c) and (d) of Fig.~\ref{fig:fig_tet_LM_CI_L_S}, one may be tempted to assign a correlation to the pnictogen substitution (using our best estimate values of $\mathrm{Pn} = (\mathrm{P}, \mathrm{As},  \mathrm{Sb})$ at $U = (0.7, 0.0, 0.8)$ eV respectively).
Notice that the absolute value of each of the contributions tends to decrease as the pnictogen site atomic number increases.
However, it is our finding that the total value of $\alpha_{xx} = - \alpha_{yy}$ does not display this correlation - as the net CI contribution seems to not follow this trend.
This case highlights the importance of tracking all of the contributions in a study of the ME response.
Focusing on our best-estimate values for the three materials, we can estimate the MEPR of each contribution.
Approximately ($190\%$, $79\%$, and $37\%$) of the total response in each of the pnictogens (P, As, Sb) respectively is due to the ${\bm \alpha}^\mathrm{CI,S}$. 
The influence of ME response from ${\bm \alpha}^\mathrm{L}$ is considerable in $\mathrm{BaMn}_2\mathrm{P}_2$ as seen in Fig.~\ref{fig:fig_tet_LM_CI_L_S}~(d) albeit with opposite sign from that of ${\bm \alpha}^\mathrm{S}$ in (c).
From this, we find that ${\bm \alpha}^\mathrm{L}$ contributes around $-143\%$ to the total value.
Finally, we should comment that ${\bm \alpha}^\mathrm{CI,L}$ amplitude in this material family is much stronger the total ${\bm \alpha}$ in other MEs probed by first principles\cite{Malashevich2012, Ricci2016, Dey2024} highlighting the fact that this term should not be neglected in a general analysis.
%


\subsection{Magnetoelectricity in the trigonal materials \label{sec:trig}}

\begin{figure}\centering
\hspace*{-8.5pt}\includegraphics[scale=0.188]{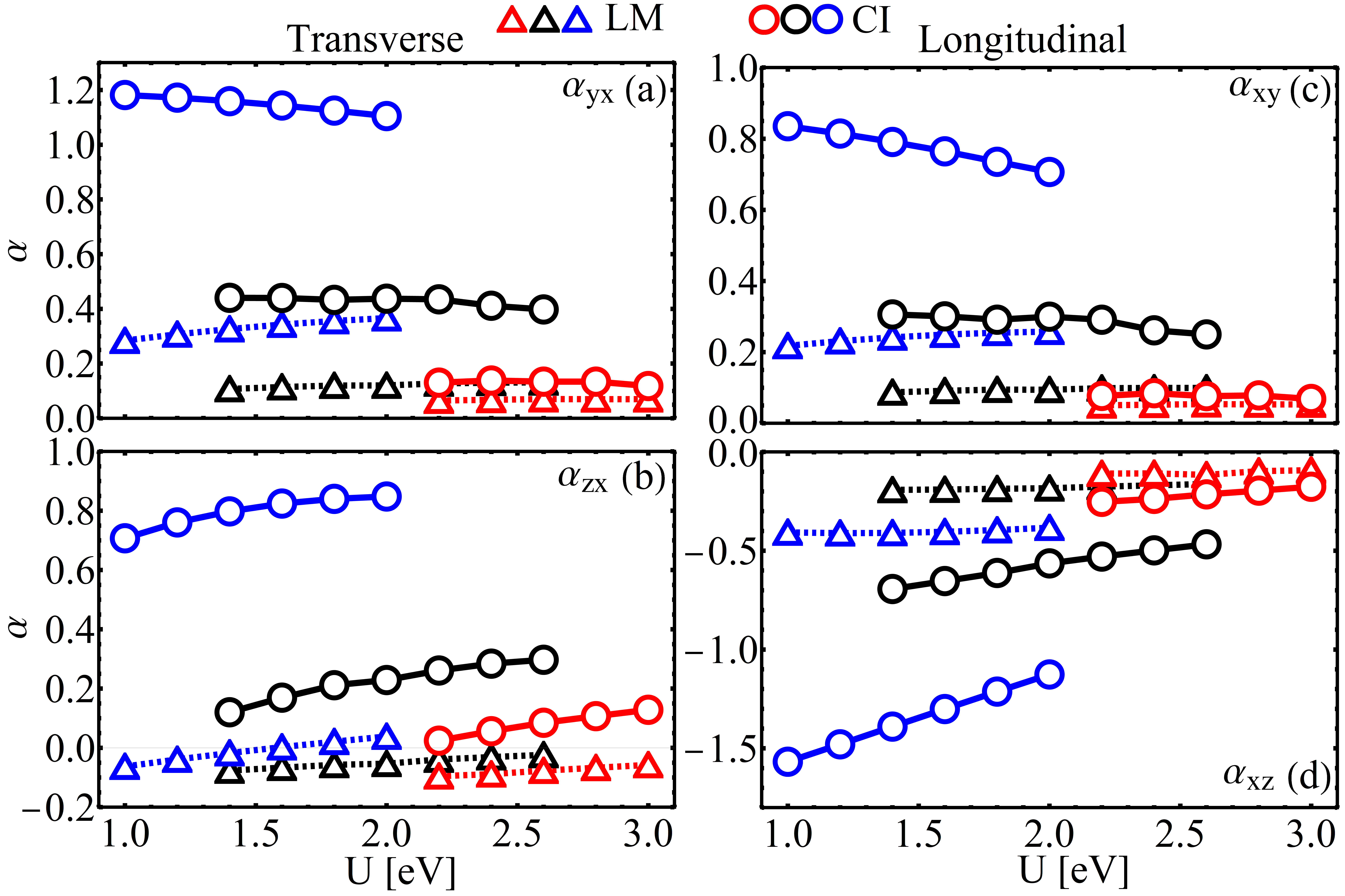}
\caption{\label{fig:fig_trig_LM_CI} Lattice-mediated (LM) in triangles and clamped-ion (CI) contribution in circles for the four (a)-(d) nonzero components of $\bm \alpha$ in $\mathrm{CaMn}_2\mathrm{Pn}$ with $\mathrm{Pn} = \mathrm{P}$ (red), $\mathrm{As}$ (black), and $\mathrm{Sb}$ (blue). Units of $\alpha_{ij}$ given in ps/m. Units of $\alpha_{ij}$ are given in ps/m.}
\end{figure}

\begin{figure}\centering
\hspace*{-8.5pt}\includegraphics[scale=0.188]{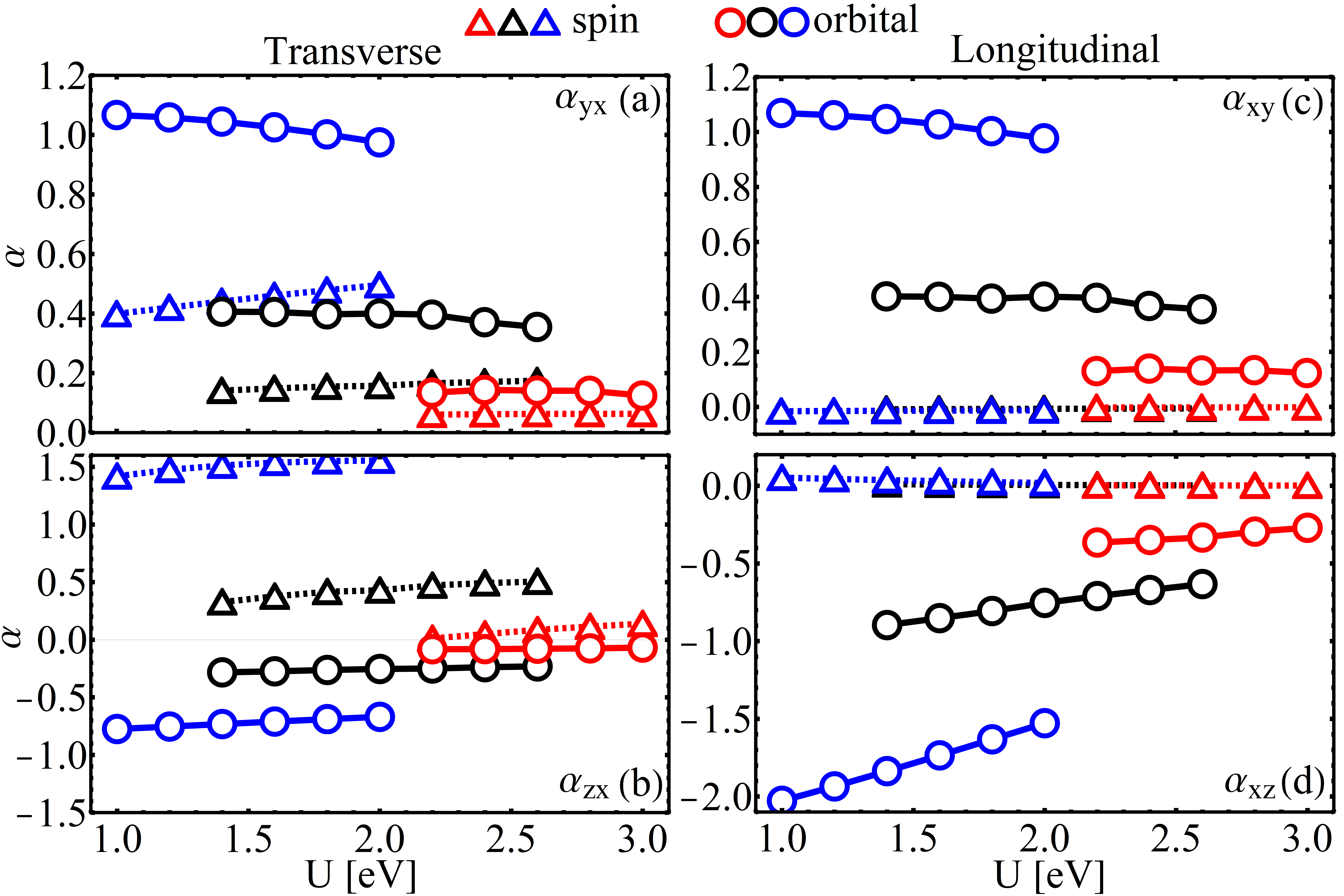}
\caption{\label{fig:fig_trig_L_S} Spin (S) in triangles and orbital (L) moment contributions in triangles for the four (a)-(d) components of ${\bm \alpha}$ in $\mathrm{CaMn}_2\mathrm{Pn}$ with $\mathrm{Pn} = \mathrm{P}$ (red), $\mathrm{As}$ (black), and $\mathrm{Sb}$ (blue). Units of $\alpha_{ij}$ are given in ps/m.}
\end{figure}

For the $P\bar{3}m1$ materials, the values of $\alpha_{yx},\alpha_{zx},\alpha_{xy}$, and $\alpha_{xz}$ are parsed into their LM (triangles) and CI contributions (circles) using Eq.~(\ref{eq:LMCI}) and are presented in panels (a), (b), (c) and (d) respectively of Fig.~\ref{fig:fig_trig_LM_CI}.
Evident in the data is a much weaker dependence on the Hubbard +U than revealed in the Ba-based compounds.
It is seen that the greatest values come from the electronic CI part (circles).
A closer inspection shows that the CI character is mixed in the $\alpha_{yx}$ and $\alpha_{zx}$ (transverse) components shown in (a) and (b).
As an example, $\mathrm{CaMn}_2\mathrm{Sb}_2$ with U = 1.6 eV (corresponding to our best estimate calculation) exhibits $\alpha_{zx}^{CI,S} = 1.36$ ps/m whereis $\alpha_{zx}^{CI,L} = -0.54$ ps/m which is a similar behavior as observed in the transverse response of the tetragonal materials in the previous section.
A similar opposite sign pattern in this component is seen upon pnictogen substitution.
In the case of the other components $\alpha_{xy}$ and $\alpha_{xz}$ in (c) and (d) which are longitudinal, we find ${\bm \alpha}^\mathrm{CI,L}$ is the primary contribution with a secondary contribution due to the orbital ${\bm \alpha}^\mathrm{LM,L}$.
In all of the systems, the involvement of the LM spin contribution is much weaker - less than $\pm 0.01$ ps/m for all components.
%
%
%
%
%
%
If we use Eq.~(\ref{eq:SL}) to instead decompose ${\bm \alpha}$, another picture forms in Fig.~\ref{fig:fig_trig_L_S}.
Here, we see the principle contribution resulting from the orbital component with again $\alpha_{yx}$ and $\alpha_{zx}$ in (a) and (b) showing a mixed character between the spin and orbital transverse response.
In (c) and (d) for the ($\alpha_{xy},\alpha_{xz}$) longitudinal components respectively, the orbital moment contribution has the largest participation in all systems with an MEPR close to $100\%$
Clearly in both representations of the data in Figs.~\ref{fig:fig_trig_LM_CI} and \ref{fig:fig_trig_L_S},
both ($\alpha_{xy},\alpha_{xz}$) and ($\alpha_{yx},\alpha_{zx}$) seem to behave differently, and the origin lies in the magnetic symmetry of the system.
With the AFM sublattice aligned along $\mathbf{x}$, the ($\alpha_{yx},\alpha_{zx}$) coefficients describe a quasi-\emph{transverse} process in the $\mathbf{y}$-$\mathbf{z}$ plane - whereis the ($\alpha_{xy},\alpha_{xz}$) depict a quasi-\emph{longitudinal} motion of the magnetization.
Note that the definitions of transverse or longitudinal are with respect to spin easy-axis.
Changes of the net spin along longitudinal directions (i.e. $\Delta M_x^\mathrm{S}$ in this case), tend to be quite stiff in AFMs.
Therefore, we are not surprised to see an essentially null spin value and a dominating orbital character as best depicted in Fig.~\ref{fig:fig_trig_L_S} (c) and (d).
Similar studies on CRO in Ref.~[\citen{Malashevich2012}] demonstrated this phenomena as well, with the changes in the magnetic spin along the longitudinal directions quite small under a field - leading to a small value of the computed $\alpha^\mathrm{S}_{zz}$. 
However, that work (as well as our benchmarking calculations presented in the Appendix) showed that the ME response due to the orbital moments in this direction could be an order of magnitude or more larger.
%
%


%
Using the analysis as in the previous section regarding the optimum value of Hubbard +U that corresponds to the ground state seen in experiment, we list the total ${\bm \alpha}$ amplitudes for the $\mathrm{(Ca,Sr)Mn}_2\mathrm{Pn}_2$ in Table~\ref{table:tab3}.
In general, we find that $\alpha_{xz} < 0$ whereis the other components of ${\bm \alpha}$ are positive (with the exception of $\alpha_{zx}$ in $\mathrm{CaMn}_2\mathrm{P}_2$.
From Table~\ref{table:tab3}, one can appreciate that ME response is much weaker than predicted in the Ba-based systems.
Still, it is interesting to see the following trend.
Upon pnictide substitution, the values of $\alpha_{ij}$ rise as the atomic number increases. 
This is contrasting to the tetragonal materials discussed in the previous section where the pnictide content depresses the total ${\bm \alpha}$ as the atomic number increases.
This alludes to the idea that the spin-orbit coupling from the ligands is much more influential in driving ${\bm \alpha}$.
Still more future investigations are warranted as to the primary driving factor in the ME susceptibility in this class of materials.
As opposed to the tetragonal materials, the coefficients are seen to not depend strongly on the +U correction with the exceptions of $\alpha_{zx}^\mathrm{CI,S}$ and $\alpha_{xy}^\mathrm{LM,S}$ which can demonstrate some sign differences depending on the Pn element albeit being small in magnitude compared to the total value. 

The substitution of Ca with Sr changes some of the response properties accordingly, but we find that the effects are minor (see Table \ref{table:tab3}).
We find remarkably indistinguishable data to that presented in Fig.~\ref{fig:fig_trig_LM_CI} and \ref{fig:fig_trig_L_S}.
Curiously, $\mathrm{SrMn}_2\mathrm{Sb}_2$ has been found to be much more insulating than its calcium cation counterpart\cite{Sangeetha2018}.
Therefore, our best estimate of ${\bm \alpha}$ uses a much larger Hubbard correction ($U = 3.8$ eV) to open the gap.
In this calculation, the MEPR for the largest coefficient $(\alpha_{zx})$ yields approximately $144\%$ and $-44\%$ due to the spin and orbital modes respectively. 
Note that a MEPR greater than 100$\%$ indicates that the other component (in this case due to orbital moments) has a negative contribution.

\begin{table}[htp!]
\caption{\label{table:tab3}%
Best estimates of the total $\alpha_{ij}$ in ps/m. The calculations are based on the ground states in Table \ref{table:tab2}.
}
\begin{ruledtabular}
\begin{tabular}{c c c c c c c c c c}
$I4/mmm$ & $\alpha_{xx}$ & $\alpha_{yy}$ & $P\bar{3}m1$ & $\alpha_{yx}$ & $\alpha_{zx}$ & $\alpha_{xy}$  &  $\alpha_{xz}$ &  & \\ 
& & & & & & & & &\\
$\mathrm{BaMn}_2\mathrm{P}_2$  & 6.41  & -6.41 & $\mathrm{CaMn}_2\mathrm{P}_2$ & 0.21 & -0.03 & 0.14 & -0.35& & \\ 
& & & & & & & & &\\ 
$\mathrm{BaMn}_2\mathrm{As}_2$  & 6.79 & -6.79 & $\mathrm{CaMn}_2\mathrm{As}_2$ & 0.55 & 0.10 & 0.39 & -0.84 &  & \\ 
& & & & & & & & &\\  
$\mathrm{BaMn}_2\mathrm{Sb}_2$  & 3.43 & -3.43 & $\mathrm{CaMn}_2\mathrm{Sb}_2$ & 1.49 & 0.83 & 1.01 & -1.71 & & \\ 
& & & & & & & & &\\ 
& & & $\mathrm{SrMn}_2\mathrm{P}_2$ & 0.19 & 0.05 & 0.13 & -0.25  & & \\ 
& & & & & & & & &\\ 
& & & $\mathrm{SrMn}_2\mathrm{As}_2$ & 0.58 & 0.16 & 0.42 & -0.76 & & \\
& & & & & & & & &\\ 
& & & $\mathrm{SrMn}_2\mathrm{Sb}_2$ & 1.19 & 1.00 & 0.621 & -0.686 & &\\ 
& & & & & & & & &\\ 
\end{tabular}
\end{ruledtabular}
\end{table}

\subsection{Hybridization effects}

The $4p$ levels in lone $\mathrm{Mn}^{+2}$ are unoccupied, but due to electronic band overlap from the pnictogen ligands, these states become partially occupied in the Zintl compounds.
As mentioned in Sec.~\ref{sec:basic}, this hybridization effect in the ground state Ba-based structures contributes significantly to $\mathbf{m}_a^\mathrm{L}$ (opposite in sign from the $d$-band contribution).
We can make a direct decomposition of Eq.~(\ref{eq:orb_contr}) separated by band occupation exploiting the fact that all contributions to ${\bm \alpha}$ are linear.
As an example, consider the following LM part due to Mn for $\mathrm{BaMn}_2\mathrm{P}_2$ (best estimate),
\begin{align}\nonumber 
{\bm \alpha}^\mathrm{LM,L} = \underbrace{\begin{pmatrix}
-0.65 & 0 & 0\\
  &  & \\
0 & 0.65 & 0\\
  &  & \\
0 & 0 & 0 \\
\end{pmatrix}}_{3d} + \underbrace{\begin{pmatrix}
-0.17 & 0 & 0\\
  &  & \\
0 & 0.17 & 0\\
  &  & \\
0 & 0 & 0 \\
\end{pmatrix}}_{4p},
\end{align}
in units of ps/m.
Similarly, the same can be done for ${\bm \alpha}^\mathrm{CI,L}$ using Eq.~(\ref{eq:orb_CIL}) where 
\begin{align}\nonumber 
{\bm \alpha}^\mathrm{CI,L} = \underbrace{\begin{pmatrix}
-8.47 & 0 & 0\\
  &  & \\
0 & 8.47 & 0\\
  &  & \\
0 & 0 & 0 \\
\end{pmatrix}}_{3d} + \underbrace{\begin{pmatrix}
-1.92 & 0 & 0\\
  &  & \\
0 & 1.92 & 0\\
  &  & \\
0 & 0 & 0 \\
\end{pmatrix}}_{4p}.
\end{align}
Here one can see that the $4p$ component cannot be neglected - and provides a significant amount of ME response as compared to the term from $3d$ electrons.
Surprisingly, ${\bm \alpha}^\mathrm{LM,L}$ and ${\bm \alpha}^\mathrm{CI,L}$ from both $3d$ and $4p$ bands have the same sign despite $\mathbf{m}_a^\mathrm{L}$ (due to the occupied $p$ states) having opposite sign to that of $d$-character.
The remainder of ${\bm \alpha}^\mathrm{CI,L}$ from the phosphorous $3p$ band contributing a magnitude of about $+0.16$ ps/m (opposite sign).
%
%
For the $P\bar{3}m1$ compounds, the $p$ band provides much less contribution to ${\bm \alpha }^\mathrm{L}$ (around $\pm 0.01$ ps/m and approximately invariant of pnictogen species).
%


%
%

\section{Discussion}

We conclude the report of our work with some comments on linear magnetoelectricity in this family and the outlook for characterizing the ME response in general.
We have provided a detailed account of how to compute the spin and orbital contributions to the linear ME tensor for both lattice-mediated and electronic (clamped-ion) parts within the DFT scheme of LDA+U.
%
%
The methodology not only captures participation ratios of the various physical processes involved but also their relative signs on equal footing.
We demonstrated good agreement with a similar full characterization of the archetypal benchmark material CRO as shown in the Appendix validating our method.
Then, we applied this effort to the family of layered AFM pnictides $\mathrm{(Ba,Ca,Sr)}\mathrm{Mn}_2\mathrm{(P,As,Sb)}_2$.
We provide basic properties information (ground state band pictures, orbital magnetization, and magnetic anisotropies) on the family of materials showing good agreement with available experimental data.
The tetragonal materials demonstrate the largest ${\bm \alpha}$ amplitudes computed.
Their predicted ME response is entirely 2D perpendicular to the collinear spin order with $\alpha_{xx} = -\alpha_{yy}$ within $\pm 10^{-3}$ ps/m precision.
The greatest value ($\alpha_{xx} \simeq 6.79$ ps/m) found in $\mathrm{BaMn}_2\mathrm{As}_2$ exceeds that of prototypical linear ME CRO by about a factor of three.
One finds that the clamped-ion spin component is the primary factor in the spin ME susceptibility.
This is not uncommon as this has been predicted in other compounds such as $\mathrm{LiNiPO}_4$ \cite{Bousquet2011} or in the narrow gap FeS \cite{Ricci2016}.
Importantly, our calculations reveal that there is an appreciable influence from the orbital magnetization along with $p$-$d$ hybridization effects in this subfamily - showing that it can give rise to non-negligible terms in both the clamped-ion and lattice-mediated orbital ME response.
A clear trend is seen upon pnictogen substitution driving the amplitudes of various contributions lower as the atomic number of the ligand atoms are increased (according to our best estimate calculations).
While the lattice-mediated and orbital contributions seem to follow this trend, the total response does not - influenced heavily by the clamped-ion terms.
While our results on the $P\bar{3}m1$ subfamily seem to give much weaker linear ME amplitudes, we showed that as the pnictogen atomic number increases, so does the amplitude of ${\bm \alpha}$.
This alludes that spin-orbit coupling from ligands are much more important than in the tetragonal materials at influencing ${\bm \alpha}$.
The largest ${\bm \alpha}$ component predicted in $\mathrm{CaMn}_2\mathrm{Sb}_2$ is then $\alpha_{xz} = -1.71$ ps/m which is comparable to the tranverse response found in CRO (see the Appendix and references therein).

In general, the procedure to compute ${\bm \alpha}$ is rather delicate. 
It depends on a number of high precision DFT calculations and is sensitive to the reciprocal space sampling, chosen structural parameters, exchange correlation functional, and force convergence criterion.
We choose to probe the technical aspect of the Hubbard+U correction in this work, providing all of our results as a function of $+U$.
This allowed us to investigate how sensitive the individual parts of ${\bm \alpha}$ are on this approximation.
While our computed ${\bm \alpha}$ for the trigonal compounds do not seem to justify a careful scan of the +U dependence, the tetragonal system certainly motivated this, demonstrating large changes despite small (relative) changes in $+U$ mainly driven by electronic effects.
It is quite apparent from our data that these electronic effects cannot be neglected and can provide substantial contributions to the ME response from both the spin and orbital modes.
The probes of ME susceptibility using first-principles methods so far have largely neglected the influence of the orbital moments. 
In the case of the prototypical CRO, this is justified by the expected quenched moments providing a very small value to the response.
However, we show an example of a materials family where this cannot not be neglected.
Thus, we emphasize that in addition to the spin (lattice-mediated) values provided by first-principles, the orbital ME response should also be characterized fully.
The notion of sign of the various contributions are also important - requiring a careful decomposition.
Still, we are optimistic that full \emph{ab initio} characterizations of the fundamental magnetoelectric interaction are useful for the ever-evolving list of applications of this class of materials.

\begin{acknowledgments}

All authors acknowledge funding from the Villum foundation Grant No. 00029378. Computations were performed on Nilfheim, a high performance computing cluster at the Technical University of Denmark. 

\end{acknowledgments}


\appendix

\section*{Appendix: benchmark case of $\mathrm{Cr}_2\mathrm{O}_3$}

\begin{figure}[b!]\centering
\includegraphics[scale=0.19]{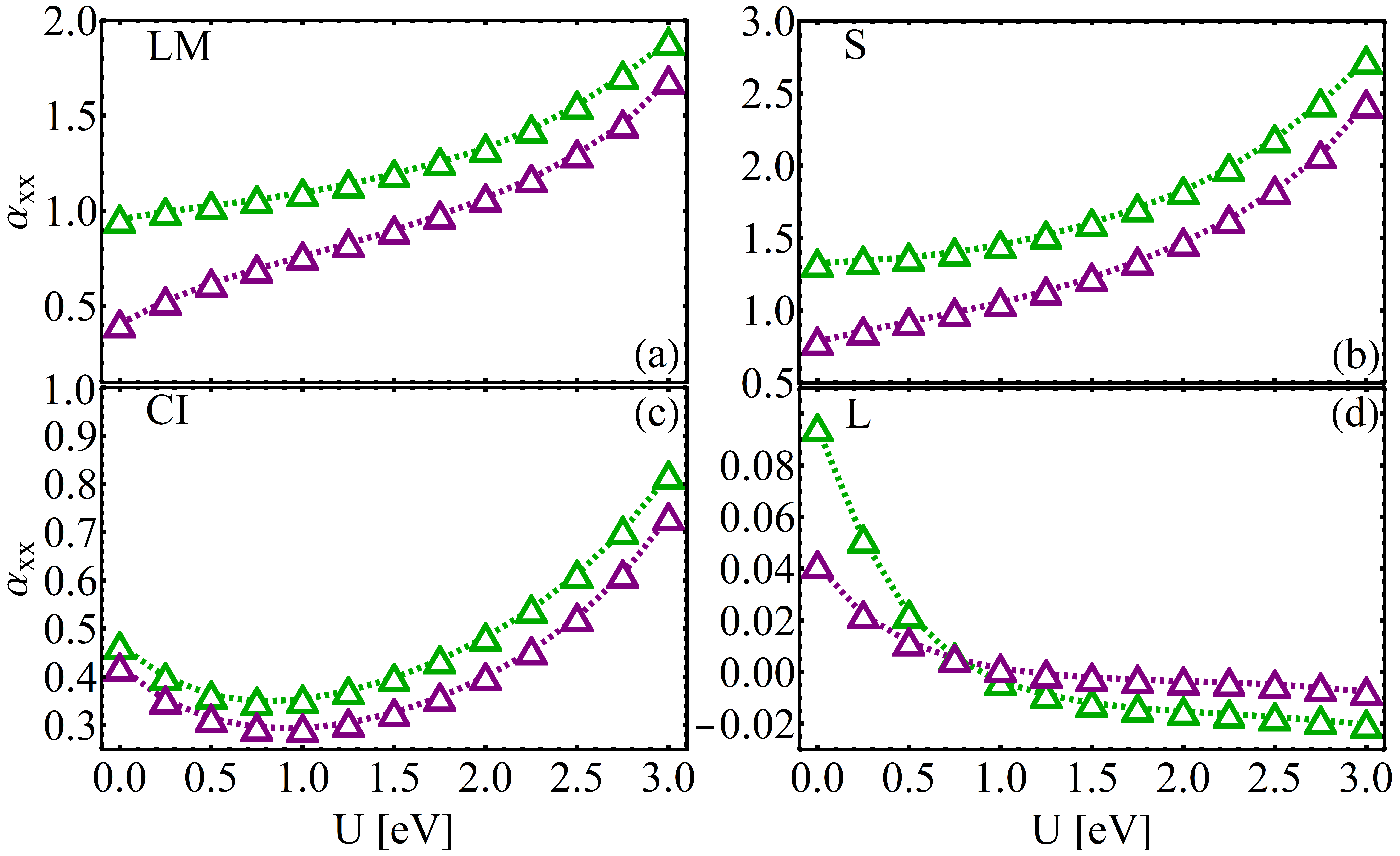}%
\caption{\label{fig:fig_CRO_trans} FLRI (green) and RLRI (purple) calculations of the transverse $\alpha_{xx}$ in ps/m for the separated LM, S, CI, and L contributions in (a) (b) (c) and (d) respectively.}
\end{figure}

As discussed in Sec.~\ref{sec:intro}, $\mathrm{Cr}_2\mathrm{O}_3$ (CRO) is a linear ME material.
It is well-studied and the richness of literature motivates us to use it as a benchmark to test and validate our method.
CRO forms as a corundum-type structure with two formula units per rhombohedral unit cell.
The $\mathrm{Cr}^\mathrm{3+}$ configuration implies 3 $\mu_\mathrm{B}$ atomic spin moments, which arrange antiferromagnetically below the Neel temperature $T_\mathrm{N} = 307$ K. 
The magnetic space group $R\bar{3}'c'$ (No. 161.106) contains the symmetry $\mathcal{I}\cdot\mathcal{T}$ (although time-reversal and inversion are separately broken) and the combination of three-fold rotational symmetry and a mirror plane containing the 3-fold axis (chosen along $z$) constrains the linear ME response to ${\bm \alpha} (R\bar{3}'c) = \mathrm{diag}(\alpha_{xx}, \alpha_{xx}, \alpha_{zz})$..
%
%
%
The sign of the ME response is determined by the AFM domain configuration \cite{Bousquet2024}, since time-reversal flips the spins by $180^\circ$ which changes the sign of all coefficients of $\alpha_{ij}$.
Using the methods discussed in Sec. \ref{sec:methods}, we study CRO at $T = 0$ K with DFT using the LDA exchange correlation functional along with the rotationally invariant Hubbard +U correction.
We include 6 valence $e^-$ for Cr ($4s^1 3d^5$) and O ($2s^2 2p^4$) using PAW cutoff radii of 1.217 \AA\ and $0.688$ \AA\ respectively).
Reciprocal space was sampled with a $6\times 6\times 6$ $\Gamma$-centered grid and we used a plane-wave cutoff of 800 eV is used.
We perform our calculations within a range of chosen +U values implemented on the $d$-orbitals of Cr to demonstrate their influence on each of the contributions to ${\bm \alpha}$.
%
%

\begin{figure}\centering
\includegraphics[scale=0.18]{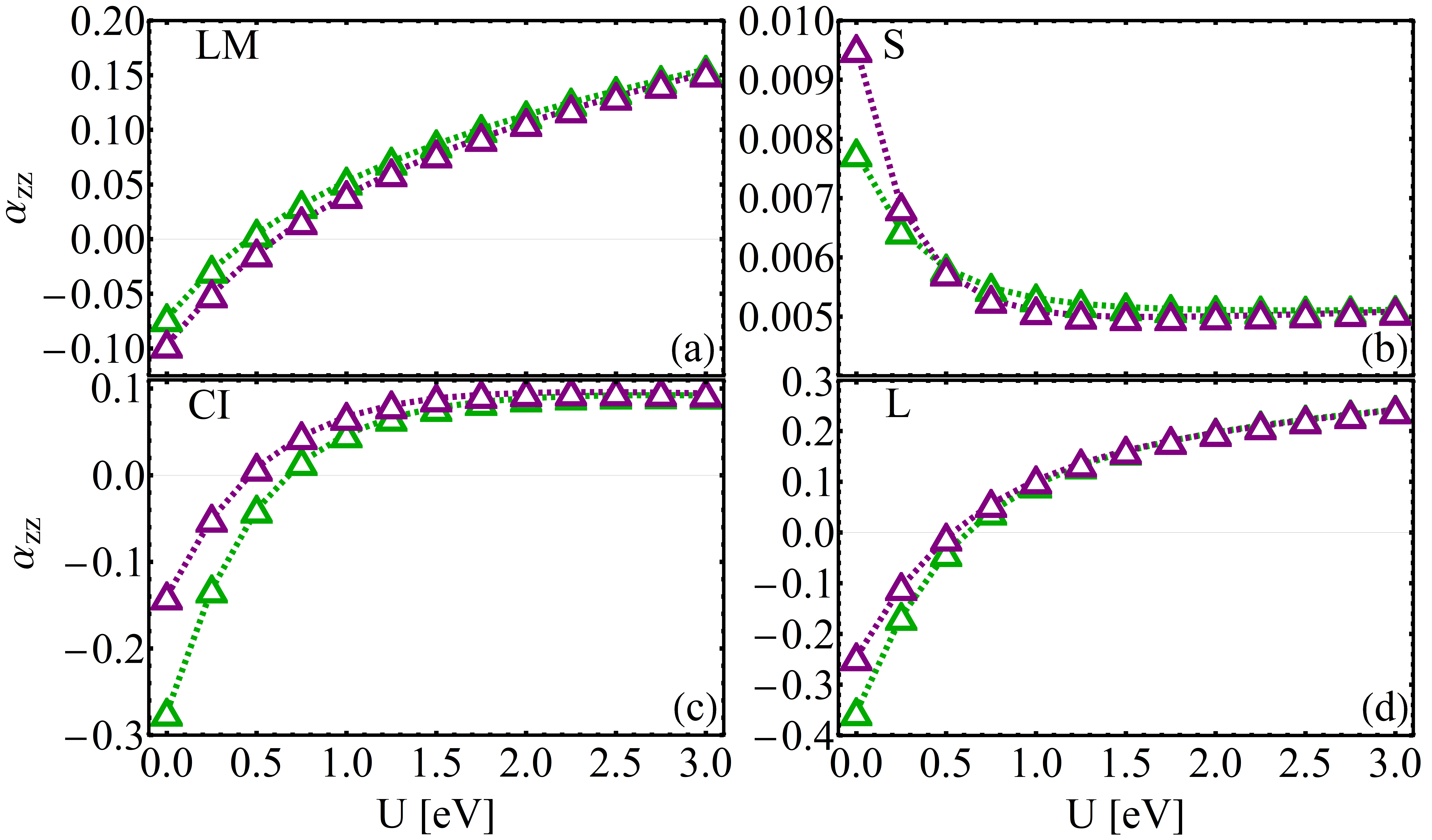}
\caption{\label{fig:fig_CRO_long} FLRI (green) and RLRI (purple) calculations of the longitudinal $\alpha_{zz}$ in ps/m for the separated LM, S, CI, and L contributions in (a) (b) (c) and (d) respectively.}
\end{figure}

We perform two different modes of calculations in the two-formula unit primitive cell.
We first fix the lattice constants to experimental values ($a = 5.358$ \AA\ and rhombohedral cell angle of $\phi = 55.0^\circ$). 
We then relax the internal coordinates and compute $\mathbf{K}$ and $Z_{j\kappa}^e$.
This mode is named as fixed-lattice relaxed-internal(FLRI).
Then, we proceed with the methods proposed in Sec. \ref{sec:methods} to compute the LM and CI responses for both the spin and orbital moments.
Next, we allow the lattice constants to relax along with the internal coordinates of the atoms - which we denote as relaxed-lattice relaxed-internal (RLRI) - and use the same approach to compute the components and contributions to ${\bm \alpha}$.
We perform both FLRI and RLRI modes as a function of the Hubbard on-site correction in a range between $0.0$-$3.0$ eV.
%
%
As for the ground state values of the nearly-quenched orbital magnetization, we find $|\mathbf{m}_a^\mathrm{L}| \simeq 0.03$ $\mu_\mathrm{B}$/Cr at U = 1.5 eV agreeing quite well with the value estimated in Ref. [\citen{Shi2009}] $|\mathbf{m}_a^\mathrm{L}| \simeq 0.04$ $\mu_\mathrm{B}$/C also using LDA.
We find only a 5$\%$ difference in $|\mathbf{m}_a^\mathrm{L}|$ between FLRI and RLRI calculations. 
%
%

Using Eqs.~(\ref{eq:LMCI}) and (\ref{eq:SL}), we parse all contributions $\mathrm{LM},\mathrm{S},\mathrm{CI},\mathrm{L}$ to the transverse component $\alpha_{xx}$ and present them in Fig.~\ref{fig:fig_CRO_trans} in (a), (b), (c), and (d) respectively.
The colors green and purple refer to FLRI and RLRI modes accordingly.
Similarly, we provide the same information for the longitudinal $\alpha_{zz}$ component in Fig.~\ref{fig:fig_CRO_long}.
While the LDA is generally known to underestimate lattice constants and cell volumes compared to experimental evidence (as we also see in the RLRI mode), one can appreciate that this does not influence our results significantly.
Instead, the main driver of changes in the values of ${\bm \alpha}$ come from the choice of the Hubbard +U.
One should note that within the range of U = $2.0\pm 0.5$ eV, the insulating gap and moments of CRO are reproduced satisfactorily. 
But even within this range, the values of various contributions can vary.

\begin{table}[t!]
\caption{\label{table:tab4}%
Calculated contributions to ${\bm \alpha}$ from the literature in ps/m as well as DFT method and code. The results of this work are presented for $U = 1.5$ eV in the FLRI mode. The range of measured amplitudes of $\alpha_{ij}^\mathrm{exp}$ from Refs. [\protect\citen{Kita1979,Wiegelmann1994}] are both performed at T$ = 4.2$ K. The same AFM domain state is considered for this comparison\cite{Bousquet2024}.
}
\begin{ruledtabular}
\begin{tabular}{c c c c c c c c c }
 Ref.: & our work & [\citen{Malashevich2012}] & [\citen{Bousquet2011}] & [\citen{Tillack2016}] & [\citen{Bousquet2024}]& \\ 
   & LDA+U & PBE & LDA+U & LDA & LDA& &\\
 & & & & & & &\\
   & GPAW\cite{Mortensen2024} & QE\cite{Giannozzi2017} & ABINIT\cite{Gonze2020} & VASP\cite{Kresse1999} & ELK\cite{elk}& &\\ 
 & & & & & & &\\
  $\alpha_{xx}$ &  & &  & & & & \\
  & & & & & & &\\
LM,S & 1.18 & 0.77 & 1.11 & 0.31 & 0.921 & & \\
& & & & & & &\\
LM,L & 0.016 & 0.025 & - & - & -&\\
& & & & & & &\\
CI,S & 0.425 & 0.22 & 0.34& 0.53 & - & & \\
& & & & & & &\\
CI,L & -0.0027 & -0.014 & -& -&- & &\\ 
& & & & & & & &\\
 $\alpha_{zz}$ &  & &  & & &  & \\ 
 & & & & & & &\\ 
LM,S & 0.003 & 0.003 & 0&  0.005 &  & & \\ 
 & & & & &  &  &\\\cline{5-7}
LM,L & 0.084 & 0.008 & - &  &  & & \\ 
& & & & $\alpha_{xx}^\mathrm{exp}$ & 0.73-1.60 & & \\ 
CI,S & 0.002 & 0.0007 & 0&  & &  &\\
& & & & $\alpha_{zz}^\mathrm{exp}$ & 0.23-0.27 &  &\\
CI,L & 0.076 & -0.009 & -& & & &\\ 
&  & & & & & &\\
\end{tabular}
\end{ruledtabular}
\end{table}

Clear trends are seen in the data, with $\alpha_{xx}^\mathrm{LM,S}$ the dominant factor in the response (not explicitly shown in Figs. \ref{fig:fig_CRO_trans} and \ref{fig:fig_CRO_long}) in agreement with previously published experimental and calculated results (see Table~\ref{table:tab4}).
Note that we only present the FLRI results at $U = 1.5$ eV in the table which is our best estimate regarding available data on the material.
The spin CI contribution factoring into an MEPR of about $36\%$ percent of the total transverse response.
The orbital MEPR ($\alpha^{L}_{xx}/\alpha_{xx}$) much lower - at only about $1\%$ percent.
%
%
For the longitudinal $\alpha_{zz}$, we find a negligible LM,S contribution (as expected since modulations along the spin easy axis are generally quite stiff in insulators).
We demonstrate that the orbital LM component is the considerable factor in driving a nonzero $\alpha_{zz}$ - although much lower than the reported low temperature measurements.
We primarily compare our results to Malashevich \emph{et} \emph{al} in Ref. [\citen{Malashevich2012}] which to our knowledge is the only work that fully characterizes ${\bm \alpha}$.
While our predictions of $\alpha_{xx}^\mathrm{LM,S}$ and $\alpha_{zz}^\mathrm{LM,L}$ are larger than those reported for calculations utilizing the Perder-Burke-Ernzerhof (PBE) exchange correlation functional, we do make note that the authors performed the LDA calculations finding larger values for ${\bm \alpha}^\mathrm{L}$ which are still comparable to our numbers within LDA+U.
%
%
We also should mention that the computed values of ${\bm \alpha}^\mathrm{L}$ are only found within the PAW spheres and therefore some contributions from interstitial regions may be missing.
Despite this, we still see good agreement of our results within the experimental range of values - and in qualitative agreement with other selected published values.
We should also note that we find the same sign and comparable value for the transverse component of ${\bm \alpha}^\mathrm{CI,L}$. 
For $\alpha^\mathrm{CI,L}_{zz}$, we do find a much larger contribution than presented in Ref. [\citen{Malashevich2012}] (but opposite sign).
%
%
However, it is hard to say if our local PAW sphere approach underpins the differences in values that we obtain or it originates in the different exchange correlation functional used.

\bibliography{apssamp}

\begin{thebibliography}{87}
\bibitem [{non()}]{nonstandNote}%
  \BibitemOpen
  \href@noop {} {\enquote {\bibinfo {title} {{The Belov-Neronova-Smirnova (BNS) magnetic space group is given in a nonstandard form. The transformation to the standard setting can be obtained by (-$\mathbf{a}$-$2\mathbf{b}, \mathbf{a}, \mathbf{c}$;0,0,0).}}}\ }\BibitemShut {NoStop}%
\bibitem [{not()}]{noteRcut}%
  \BibitemOpen
  \href@noop {} {\enquote {\bibinfo {title} {{We set the cut-off radii (in \AA) of the PAW spheres as follows: {Ba}($1.16$), {Sr}($1.29$), {Ca}($1.17$), {P}(0.95), {As}(1.06), and {Sb}(1.23) as implemented in the default atomic PAW potential settings of {GPAW} commit version ce976839bd3631d9bab285a2ff27476b2b6b46e0.}}}\ }\BibitemShut {NoStop}%
\end{thebibliography}

\end{document}